\documentclass[11pt]{article}

\usepackage{amsfonts,amssymb,cellspace,chet,colortbl,dsfont,enumerate,graphicx,mathrsfs,mathtools,multicol,subcaption,tikz}

\newcommand{\rep}[1]{\ensuremath{\mathbf{#1}}}

\newcommand{\one}{\ensuremath{\mathbf{1}}}

\newcommand{\an}[1][n]{\ensuremath{\mathfrak{a}_{#1}}}
\newcommand{\bn}[1][n]{\ensuremath{\mathfrak{b}_{#1}}}
\newcommand{\cn}[1][n]{\ensuremath{\mathfrak{c}_{#1}}}
\newcommand{\dn}[1][n]{\ensuremath{\mathfrak{d}_{#1}}}
\newcommand{\g}{\ensuremath{\mathfrak{g}}}
\newcommand{\h}{\ensuremath{\mathfrak{h}}}

\newcommand{\ml}[1]{\begin{tabular}{@{}c@{}} #1 \end{tabular}}
\newcommand{\pr}[1]{\left(#1\right)}

\newcommand{\cb}[1]{\left\{#1\right\}}
\newcommand{\floor}[1]{\left\lfloor #1 \right\rfloor}
\newcommand{\wt}{\widetilde}
\newcommand{\fa}{\, \forall \:} 
\newcommand{\tr}[1]{\text{tr}\hspace{-2pt}\pr{#1}} 

\title{A No-Go Theorem for Fully SGUTs\\with Metastable SUSY Breaking}

\author{Jean-Fran\c{c}ois Fortin\email{jean-francois.fortin@phy.ulaval.ca} and Jean-Samuel Leboeuf\email{jean-samuel.leboeuf.1@ulaval.ca}}

\affiliation{
D\'epartement de Physique, de G\'enie Physique et d'Optique,\\Universit\'e Laval, Qu\'ebec, QC G1V 0A6, Canada
}

\abstract{We introduce fully SGUTs, SUSY grand unified theories that, upon symmetry breaking through the Higgs mechanism, decompose into a visible sector and an extra sector where the dynamics of the extra sector gauge group is responsible for SUSY breaking.  Fully SGUTs thus have the important feature that all gauge groups of the visible sector and the extra sector unify into a simple gauge group at the SGUT scale, therefore generalizing the successful MSSM gauge coupling unification to all the gauge couplings of the theory.  By focusing on the ISS SUSY-breaking mechanism in the extra sector, we show that it is impossible to reproduce the MSSM matter content when there exists a metastable ISS SUSY-breaking state.}

\date{June 2017} 

\begin{document}

\maketitle



\section{Introduction}\label{Sec:Intro}

As of this writing, the Standard Model (SM) of particle physics seems to explain all physics up to the energy scale probed by the Large Hadron Collider (LHC).  Experimental observations and theoretical arguments suggest however that the SM is not the full story.  For example, the SM cannot account for neutrino masses and it does not have a suitable dark matter candidate.  Moreover, the electroweak scale seems highly fine-tuned unless an appropriate mechanism is found to explain its smallness compared to the Planck scale.  One such mechanism is supersymmetry (SUSY) \cite{Ramond:1971gb,Neveu:1971rx,Gervais:1971ji,Golfand:1971iw,Volkov:1973ix,Wess:1974tw}, where the introduction of SUSic partners, or superpartners, stabilizes the electroweak scale \cite{Dimopoulos:1981au,Dimopoulos:1981zb,Sakai:1981gr,Kaul:1981hi}.  SUSY must however be spontaneously broken in our Universe and for the electroweak scale to be naturally small, superpartners must be relatively light, in apparent tension with their non-observation at the LHC \cite{Buckley:2016tbs}.  Although the LHC puts strict constraints on the superpartner spectrum, somewhat discrediting SUSY as a solution to the hierarchy problem, with its theoretical consistency SUSY is still one of the best theories of beyond-the-SM physics.

Actually, one of the most intriguing properties of the minimal SUSic SM (MSSM) \cite{Fayet:1976et,Fayet:1977yc,Farrar:1978xj,Fayet:1979sa} is gauge coupling unification \cite{Ellis:1990wk,Amaldi:1991cn,Langacker:1991an,Giunti:1991ta}, suggesting a SUSic grand unified theory (SGUT) \cite{Buras:1977yy,Dimopoulos:1981zb,Sakai:1981gr}.  Indeed, that the extra SUSic partners contribute to the MSSM gauge coupling $\beta$-functions in such a way that the three gauge couplings meet relatively well at the SGUT scale $\Lambda_\text{SGUT}\approx10^{16}\,\text{GeV}$ points towards the possibility that the three MSSM gauge groups merge into a larger gauge group, like $SU(5)$ or $SO(10)$, leading to a SGUT.

SUSY must also be spontaneously broken and there are several avenues to accomplish this.  One interesting possibility is to introduce a SUSY-breaking sector with an extra gauge group $G_\text{SB}$ and extra matter fields where SUSY is spontaneously broken dynamically \cite{Dine:1981za,Witten:1981nf,Dine:1993yw,Dine:1994vc,Dine:1995ag,ArkaniHamed:1997jv,Giudice:1998bp}.  The breaking is then gauge-mediated to the MSSM sector with the help of some messenger fields, \textit{i.e.} SUSY-breaking sector fields charged under the MSSM gauge group.  In this framework, it seems odd that gauge coupling unification occurs for the MSSM gauge couplings but not necessarily for the extra $G_\text{SB}$ gauge coupling, leading to a partial (instead of a complete) unification of the forces of nature.

In this paper we investigate the idea that an extra sector with gauge group $G_\text{SB}$, the dynamics of which are responsible for spontaneous SUSY breaking, unifies with the MSSM gauge group $G_\text{SM}$, leading to what we call a fully SGUT with simple gauge group $G_\text{SGUT}\supset G_\text{SB}\times G_\text{SM}$.  In order to obtain a consistent fully SGUT, several problems must be properly addressed.  First, it is necessary that the spontaneous breaking of the simple gauge group $G_\text{SGUT}$ into its subgroups $G_\text{SM}\times G_\text{SB}$ can occur through a given mechanism, like the Higgs mechanism with an appropriate scalar potential, and not all symmetry breaking patterns are guaranteed to occur \cite{Li1974,Elias1975comment}.  It is also required that the (anomaly-free) matter content in the fully SGUT breaks into the MSSM matter content as well as appropriate matter fields in the extra sector.  The latter must include fields that lead to spontaneous SUSY breaking as well as fields that play the role of the messenger fields mediating SUSY breaking to the MSSM.  Then, from the bottom-up perspective it is necessary that all the gauge groups unify at the SGUT scale, which is highly dependent on the matter content of the theory.  From the top-down point of view, this puts some constraints on the dynamics of the extra sector, which must nevertheless exhibit spontaneous SUSY breaking.  Finally, although there is not much renormalization group flow running from the SGUT scale to the Planck scale, it would be preferable that the gauge group $G_\text{SGUT}$ is asymptotically free to avoid Landau poles, and this is again dependent on the matter content of the theory.

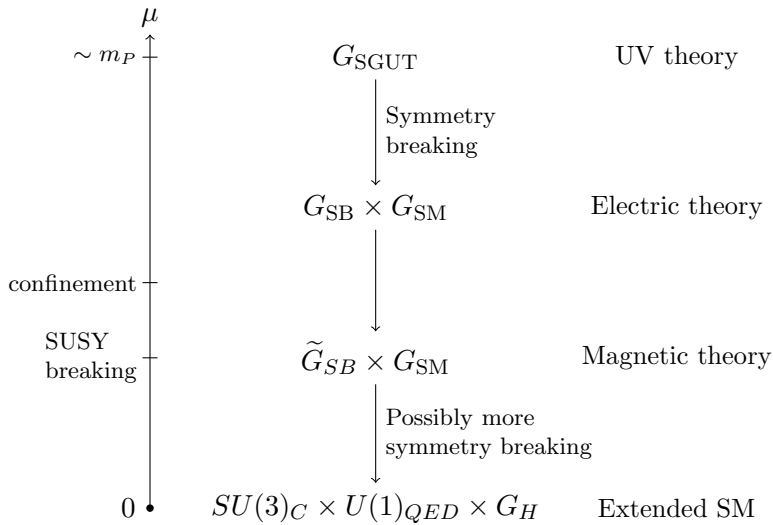
\begin{figure}[t!]
\begin{center}
\newcommand{\fs}[1]{\fontsize{#1}{144}\selectfont}
\begin{tikzpicture}[every text node part/.style={align=left}]
	
	\path 	(0,6) node[](G_SGUT){$G_\text{SGUT}$} --
			(0,4) node[](GSB){$G_\text{SB}\times G_\text{SM}$} -- 
			(0,2) node[](GSBtilda){$\wt{G}_{SB}\times G_\text{SM}$} --
			(0,0) node[](SU3){$SU(3)_C \times U(1)_{QED} \times G_H$};

	\draw[->] (G_SGUT) -- node[pos=.5,right](SB1){\footnotesize Symmetry \\[-6pt]\footnotesize breaking} (GSB);
	\draw[->] (GSB) -- (GSBtilda);
	\draw[->] (GSBtilda) -- node[right](SB2){\footnotesize Possibly more \\[-6pt]\footnotesize  symmetry breaking} (SU3);

	\draw[->] (-3,0) node[fill=black,circle,inner sep=1pt,label={left:$0$}](origin){}-- (-3,6.3) coordinate(axis);
	\node[anchor=-90,fill=white,outer sep=2,inner sep=1](mu) at (axis) {$\mu$};
	\draw (-3.1,6) -- (-2.9,6) node(mP)[left, xshift=-4pt]{\footnotesize $\sim m_P$};
	\draw (-3.1,2) -- (-2.9,2) node(SUSYB)[left, xshift=-4pt]{\footnotesize SUSY \\[-6pt] \footnotesize  breaking};
	\draw (-3.1,3) -- (-2.9,3) node(confinement)[left, xshift=-4pt]{\footnotesize confinement};

	\node[font=\fs{10}](UV) at (4,6) {UV theory};
	\node[font=\fs{10}](electric) at (4,4) {Electric theory};
	\node[font=\fs{10}](magnetic) at (4,2) {Magnetic theory};
	\node[font=\fs{10}](ESM) at (4,0) {Extended SM};

\end{tikzpicture}	
\end{center}

\caption{Sketch of the fully SGUT framework with metastable SUSY breaking.}
\label{fig:sketch}
\end{figure}

Before attacking these problems, we must focus on a particular implementation of fully SGUTs to make it more tractable.  More specifically, we introduce the idea of fully SGUT, where all gauge groups unify at the SGUT scale $\Lambda_\text{SGUT}$, with the extra gauge group of the extra sector being $SU(N_c)$, $SO(N_c)$ or $Sp(2N_c)$.  Moreover, the extra gauge group is responsible for spontaneous SUSY breaking through a metastable state \textit{\`a la} Intriligator, Seiberg and Shih (ISS) \cite{Intriligator:2006dd}, arguably one of the easiest ways to spontaneously break SUSY.  Finally, to allow for complex representations, the fully SGUT gauge group is chosen to be $SU(N)$ or $SO(N)$.\footnote{Complex representations exist also for the exceptional Lie algebra $E_6$, but it is clearly too small for our purpose.}  A sketch of our framework is shown in Figure~\ref{fig:sketch}.

Hence the fully SGUT gauge group must be $G_\text{SGUT}=SU(N)$ or $G_\text{SGUT}=SO(N)$, the gauge group $G_\text{SM}$ must have rank at least $4$ to accommodate the MSSM, and the irreducible representations of the matter fields in $G_\text{SGUT}$ must at the very least result in the MSSM matter content (the three generations and the Higgs sector) upon spontaneous symmetry breaking.  This last statement can be stated concisely in mathematical terms since the decomposition of an irreducible representation $R$ in $G_\text{SGUT}$ into the subgroup $G_\text{SB}\times G_\text{SM}$ can be written as
\begin{align}\label{eq:Irrep}
R\downarrow\bigoplus_i m_i\,r^{G_\text{SB}}_i\times r^{G_\text{SM}}_i,
\end{align}
where $r_i^G$ is an irreducible representation of $G$ and $m_i$ is the multiplicity.  We are thus looking for irreducible representations $R$ that contain in their decomposition the MSSM matter content.  Note that it would be more appropriate here for the irreducible representations that contain the MSSM particles to be singlets under the extra gauge group that ultimately breaks SUSY.  Indeed, upon confinement charged fields usually acquire a mass of the order of the confinement scale, apart from the Goldstone bosons (pseudo-Goldstone bosons) which stay massless (light), and therefore cannot play the role of the MSSM matter fields.  More importantly, too much matter fields from the point of view of the extra gauge group $G_\text{SB}$ could render it infrared free instead of asymptotically free, forbidding the existence of the SUSY-breaking vacuum.  Hence the MSSM matter content should not be charged under the extra gauge group.

For the metastable SUSY-breaking vacuum to reliably exist, it is necessary that the extra gauge group $G_\text{SB}$ is asymptotically free and that the dual magnetic theory is infrared free to exhibit the SUSY-breaking minimum.  The first condition implies that the one-loop $\beta$-function contribution $b$ to the extra gauge coupling is larger than zero.  This translates into
\begin{itemize}
\item $b=3N_c-N_f>0$ for $G_\text{SB}=SU(N_c)$,
\item $b=3(N_c-2)-N_f>0$ for $G_\text{SB}=SO(N_c)$,
\item $b=3(N_c+1)-N_f>0$ for $G_\text{SB}=Sp(2N_c)$,
\end{itemize}
where $N_c$ (and $N_c-2$ and $N_c+1$ for different Lie groups) is the number of colors given by the quadratic Casimir $C_2$ of the adjoint representation and $N_f$ is the generalized number of flavors to be defined shortly.  The second condition demands that the dual magnetic theory is in the free magnetic range \cite{Intriligator:2006dd}, which leads to
\begin{itemize}
\item $N_c+1\leq N_f<\frac{3}{2}N_c$ for $G_\text{SB}=SU(N_c)$,\hfill\refstepcounter{equation}\textup{(\theequation)\label{eq:ISS_SU(N)}}%
\item $N_c-2\leq N_f<\frac{3}{2}(N_c-2)$ for $G_\text{SB}=SO(N_c)$,\hfill\refstepcounter{equation}\textup{(\theequation)\label{eq:ISS_SO(N)}}%
\item $N_c+1\leq N_f<\frac{3}{2}(N_c+1)$ for $G_\text{SB}=Sp(2N_c)$,\hfill\refstepcounter{equation}\textup{(\theequation)\label{eq:ISS_Sp(2N)}}%
\end{itemize}
such that the first condition is trivially satisfied when the second criterion is verified.  Here, the generalized number of flavors $N_f$ is defined as the sum of the Casimirs $C(r)$ for each irreducible representation $r$ in $G_\text{SB}$.  It can be computed explicitly from the SGUT gauge group matter content using the formalism of \eqref{eq:Irrep} as
\begin{align}\label{eq:Nf_natural_def}
N_f = \sum_R\sum_im_i\text{dim}(r_i^{G_\text{SM}})C(r_i^{G_\text{SB}})+\cdots, 
\end{align}
where $\text{dim}(r_i^{G_\text{SM}})$ is the dimension of the irreducible representation $r_i^{G_\text{SM}}$.  Due to their large mass, the massive gauge bosons generated through the Higgs mechanism are not included in the generalized number of flavors, hence the ellipses.  It is also assumed that all the irreducible representations $r_i^{G_\text{SB}}$ are light and thus contribute to the generalized number of flavors.

In this paper, we will show that fully SGUTs are not possible in the specific framework described above.  Indeed, apart from all the problems one has to address before studying the phenomenology of such models, successfully implementing fully SGUTs with metastable SUSY breaking requires that the matter content of the extra sector leads to a dual magnetic theory in the free infrared range, the appropriate window of colors and flavors necessary for the existence of the metastable SUSY-breaking state.  Therefore, even before we can address the points mentioned above, we demonstrate that it is not possible to have an ISS metastable SUSY-breaking state in the extra sector while having the MSSM matter content in the visible sector.

We prove the no-go theorem with the help of several relations between the Casimirs of different irreducible representations in $G_\text{SGUT}$ and thus find all irreducible representations which satisfy the condition for the existence of a metastable SUSY-breaking vacuum.  We finally show that the MSSM matter content cannot be recovered from these irreducible representations.  Specifically, the allowed irreducible representations lead to specific symmetry breaking patterns for which the branching rules cannot accommodate three SM families.

Hence we show that for an appropriate number of generalized flavors $N_f$ such that a SUSY-breaking minimum appears through the ISS mechanism, it is not possible to get the MSSM matter content for any of the possible symmetry breaking patterns permitted by the allowed irreducible representations.  Our no-go theorem is therefore general, being based only on the generalized number of flavors (assuming all matter fields generated by the spontaneous symmetry breaking are light), the rank of the SM, and assumption that the symmetry breaking of $G_\text{SGUT}$ occurs through the Higgs mechanism where a Higgs field acquires a vacuum expectation value (VEV).

In view of the previous observations, we conclude that the specific framework of fully SGUTs with metastable SUSY breaking \textit{\`a la} ISS cannot occur.  It would be interesting to find a successful example of fully SGUT using model-building techniques to give large masses to some unwanted matter fields or using another mechanism for spontaneous SUSY breaking.  The phenomenology of such models could give valuable insights on the superpartner spectrum and maybe shed some light on the hierarchy problem and/or the reason why superpartners have not been observed at the LHC yet (if nature is SUSic).

This paper is organized as follows: Section~\ref{sec:preliminaries} proves an identity between the generalized number of flavors for an irreducible representation $R$ of $G_\text{SGUT}$ and the Casimir of $R$ in the fully SGUT gauge group.  To obtain the complete set $\{R\}$ of irreducible representations that allow for the ISS SUSY-breaking mechanism to occur, another relation between the Casimirs of different irreducible representations in $G_\text{SGUT}$ is also proven.  Section~\ref{sec:no-go_theorem} then demonstrates the no-go theorem for $G_\text{SGUT}=SU(N)$ and $G_\text{SGUT}=SO(N)$ gauge groups.  Section~\ref{sec:conclusion} concludes with a discussion of the no-go theorem while some of the computations are shown in appendix~\ref{app:lie_algebras}.


\section{Preliminaries}\label{sec:preliminaries}

In this section we prove an identity between the generalized number of flavors and the Casimir of the irreducible representations.  By studying chains of maximal subgroups, we then determine the less restrictive constraint on the generalized number of flavors.  We also introduce an ordering in function of the Casimirs which will be helpful in restricting the irreducible representations allowed by our framework.

\subsection{Considerations on the ISS conditions}\label{sec:ISS}

The ISS conditions put strong constraints on the matter field irreducible representations allowed in our framework.  Since different SGUT gauge group symmetry breaking patterns lead to different ISS conditions, it is necessary to understand what happens to the ISS conditions under symmetry breaking.  To cover all possibilities allowed by our framework, it is important to determine the less stringent upper bound on the generalized number of flavors to ensure that all irreducible representations permitted by the ISS conditions are studied.  Indeed, it will be shown that the ISS upper bounds become more stringent after symmetry breaking.

The two quantities of interest in the ISS conditions \eqref{eq:ISS_SU(N)}, \eqref{eq:ISS_SO(N)} and \eqref{eq:ISS_Sp(2N)}, $N_f$ and $N_c$, are related to the Casimir $C$ and the quadratic Casimir $C_2$ of their relevant group.  There exist several equivalent definitions of the Casimirs for an irreducible representation $R$ of a given Lie group $G$.  For example, from the definitions \cite{Peskin1995} we obtain
\begin{align}
\tr{t^A_Rt^B_R}=C(R)\delta^{AB},\quad\quad t^A_Rt^A_R=C_2(R)\cdot\mathds{1},\label{eq:casimir_def}
\end{align}
where the $t^A_R$ are the generators of the group $G$ in the irreducible representation $R$.  The set $\left\{t^A\right\}$ forms a basis for the Lie algebra $\g$ of $G$.  The Casimir $C$ provides a normalization of the generators such that once it is set for a particular irreducible representation, all the Casimirs of the other irreducible representations are fixed.  For convenience, we also use $\g$ to express the adjoint representation of $G$.

From these group theoretic quantities, we can uniquely express an upper bound on the generalized number of flavors (without symmetry breaking) as
\begin{align}\label{eq:ISS}
N_f <\frac{3}{2}C_2(\g_\text{SB}),
\end{align}
where $C_2(\g_\text{SB}) = N_c$ for $SU(N_c)$, $N_c-2$ for $SO(N_c)$ and $N_c+1$ for $Sp(2N_c)$.  The three ISS conditions \eqref{eq:ISS_SU(N)}, \eqref{eq:ISS_SO(N)} and \eqref{eq:ISS_Sp(2N)} are thus simultaneously satisfied by this inequality.  It is clear from \eqref{eq:casimir_def} that the quadratic Casimir of the adjoint representation is equal to the Casimir of the adjoint representation.  Hence the common ISS condition \eqref{eq:ISS} can be re-expressed solely in terms of Casimirs $C$ as
\begin{align}\label{eq:ISSC}
N_f<\frac{3}{2}C(\g_\text{SB}).
\end{align}
This form of the ISS condition is the most convenient to study the effects of symmetry breaking.

Since any subgroup of a given classical Lie group falls within a chain of relatively maximal subgroups, it is only necessary to study maximal subgroups of general Lie groups.  Although the complete list of maximal subgroups will not be needed, we nevertheless present all maximal subgroups of classical Lie groups, knowing that more elaborated subgroups, and hence any symmetry breaking pattern, can be achieved by using maximal subgroups recursively.  To ensure all possible irreducible representations are taken into account, we find the largest upper bound on $N_f$ \eqref{eq:Nf_natural_def} allowed by the ISS condition \eqref{eq:ISSC} from these maximal subgroups.

All maximal subgroups of classical Lie groups were found by Dynkin \cite{dynkin1952maximal}.  They form two categories, the non-simple and the simple subgroups, which will be denoted as embeddings of the first and second kind respectively.

The non-simple ones are exhausted by the following list, where regular and special embeddings are included with no differentiation:
\begin{multicols}{2}
\begin{enumerate}[(i)]
\setlength\itemsep{0pt}
\item $SU(N)\downarrow SU(N-M)\times SU(M)\times U(1)$,
\item $SU(NM)\downarrow SU(N)\times SU(M)$,
\item $SO(N)\downarrow SO(N-M)\times SO(M)$,\label{lst:SO_embedding_in_SO}
\item $SO(NM)\downarrow SO(N)\times SO(M)$,
\item $SO(N)\downarrow SU(\floor{N/2})\times U(1)$,
\item $SO(4MN)\downarrow Sp(2N)\times Sp(2M)$,
\item $Sp(2N)\downarrow Sp(2N-2M)\times Sp(2M)$,
\item $Sp(2MN)\downarrow Sp(2N)\times SO(M)$,
\item $Sp(2N)\downarrow SU(N)\times U(1)$.
\end{enumerate}
\end{multicols}
The second category corresponds to special simple subgroups embedded in the following way: For any irreducible representation $R_*$ of a group $G$, classical or exceptional, with dimension $d_*$, we have that $G$ is a subgroup in $SU(d_*)$.  However this embedding may not be maximal.  If $R_*$ is real, then $G$ is a maximal subgroup of $SO(d_*)$, while if it is pseudoreal, $G$ is maximal in $Sp(d_*)$.  For some very specific cases, $G$ may be not maximal, but this is not relevant in the following.

Before proceeding, we make a point here about exceptional Lie groups.  Exceptional groups can also be simply embedded in classical groups.  However, they do not have dual groups which could break SUSY.  Furthermore, we can consider their maximal classical subgroups as non-maximal subgroups of their embedding group.  In this way, we can eliminate exceptional groups from our present analysis.

Intuitively, the most direct way to solve the problem at hand would be to look for irreducible representations in each classical Lie groups that can generate the MSSM content upon symmetry breaking.  For this, we would need to find the branching rules of these irreducible representations, which describe how irreducible representations decompose \cite{whippman1965branching,thoma2000weyl}.  Branching rules are however far from trivial or even easy to compute in general since they are different for every group and symmetry breaking pattern.  With these rules, we could try to find which irreducible representations could contain the MSSM.  Then, we would compute the generalized number of flavors $N_f$ to eventually conclude that, in most cases, they do not satisfy the ISS condition \eqref{eq:ISSC}.

There is a better way to approach the problem, by making the enlightening observation that the quantity $\sum_im_i\dim(r_i^{G_\text{SM}})C(r_i^{G_\text{SB}})$ of interest in the computation of the generalized number of flavors is given by the Casimir $C(R)$ (up to a constant) of the original irreducible representation $R$ in $G_\text{SGUT}$, \textit{i.e.}
\begin{align}
\sum_im_i\dim(r_i^{G_\text{SM}})C(r_i^{G_\text{SB}})=\xi^2C(R).\label{eq:Nf_simplified_eq}
\end{align}
The constant $\xi$, which is necessary to properly rescale the generators, is $1$ for embeddings of the first kind and $\sqrt{C(r_*)/C(R_*)}$ for embeddings of the second kind.

To prove \eqref{eq:Nf_simplified_eq}, we now consider a subgroup $H$ of $G$.  The subgroup $H$ has for Lie algebra $\mathfrak{h}$, a subalgebra of $\g$.  We can choose a basis for $\mathfrak{h}$ as a subset of the $t^A$ with exactly $\dim\mathfrak{h}$ generators.  When the symmetry is reduced from $G$ to $H$, the irreducible representation $R$ of $G$ breaks as
\begin{align}
G&\downarrow H\nonumber\\
R&\downarrow r\equiv\bigoplus_im_ir_i,\label{eq:r_decomposition}
\end{align}
where $r$ is a representation of $H$ that may be reducible.  This representation can be block-diagonalized as a direct sum of irreducible representations $r_i$ of $H$.  Here $m_i$ stands for the multiplicity of each such irreducible representation.

To better understand what happens to the Casimir upon symmetry breaking, it is convenient to use the exponential map development of the elements of $G$.  Indeed, an element $g\in G$ in an irreducible representation $R$, near the identity, can be parametrized by a set $\cb{\alpha^A}$ as
\begin{align*}
g=\exp\pr{\sum_{A=1}^{\dim \g}\alpha^At^A_R}.
\end{align*}
Under symmetry reduction $G\downarrow H$, only some (properly-normalized) linear combinations of the generators are kept.  The resulting elements of $G$ that belongs to $H$ can thus be written as
\begin{align*}
h=\exp\pr{\sum_{a=1}^{\dim \mathfrak{h}}\beta^at^a_r},
\end{align*}
where the set of $\cb{t^a_r=\xi U^{aA}t^A_R}$ forms a complete basis for $H$.  Here $\xi$ properly normalizes the generators as above and $U$ singles out the proper linear combinations.  The rectangular matrices $U$ are normalized such that $U^{aA}U^{bB}\delta^{AB}=\delta^{ab}$ and satisfy $\xi U^{aA}U^{bB}f^{ABC}=f^{abc}U^{cC}$ where the $f$'s are the appropriate structure constants.

This implies that the generators of $H$ in representation $r$ are related to the generators of $G$ in representation $R$.  Hence, according to \eqref{eq:casimir_def}, we have
\begin{align}
C(r)=\tr{t^a_rt^a_r}=\xi^2U^{aA}U^{aB}\tr{t^A_Rt^B_R}=\xi^2C(R)\quad\text{(no sum on $a$)}.\label{eq:casimir_equality}
\end{align}
On the other hand, it is possible to block-diagonalize the $\cb{t_r^a}$ so that each block is an irreducible representation $r_i$ of $H$ according to the decomposition \eqref{eq:r_decomposition}.  Then the trace can be taken on each block separately, which leads to
\begin{align}
C(r)=\sum_im_i\tr{t^a_{r_i}t^a_{r_i}}=\sum_im_iC(r_i)\quad\text{(no sum on $a$)}.\label{eq:casimir_rep_to_irreducible representation}
\end{align}

We are interested in the symmetry breaking $G_\text{SGUT}\downarrow G_\text{SB}\times G_\text{SM}$, so in our case $H$ is the direct product of two disjoint subgroups (by that we mean that the only element intersecting both is the identity).  Replacing $H$ by $H\times H'$ in the previous analysis alternatively yields
\begin{align}
G&\downarrow H\times H'\nonumber\\
R&\downarrow r=\bigoplus_im_i\pr{r_i\times r'_i},\label{eq:r_decomposition_product_group}
\end{align}
where $r_i\times r_i'$ is an irreducible representation of $H \times H'$ made from the direct product of an irreducible representation $r_i$ of $H$ and of an irreducible representation $r_i'$ of $H'$.

To achieve the desired result, we further reduce the symmetry by taking only $H$ as a subgroup of $H\times H'$.  This is equivalent to replacing $r_i'$ in \eqref{eq:r_decomposition_product_group} by a sum of $\dim\pr{r_i'}$ identity representations.  Mathematically, we can write this as
\begin{align*}
\begin{array}{r c c c l}
G &\downarrow & H\times H' & \downarrow & H\\
R &\downarrow & r & \downarrow & \displaystyle\bigoplus_im_i\dim\pr{r_i'}r_i.
\end{array}
\end{align*}
From this expression, and from \eqref{eq:casimir_equality} and \eqref{eq:casimir_rep_to_irreducible representation}, we get the following identity
\begin{align}
\sum_im_i\dim\pr{r_i'}C(r_i)=\xi^2C(R),\label{eq:id_casimir}
\end{align}
which is exactly the internal sum of the definition of $N_f$ \eqref{eq:Nf_natural_def} with $H=G_\text{SB}$ and $H'=G_\text{SM}$, thus proving \eqref{eq:Nf_simplified_eq}.

This result simplifies greatly our analysis.  Indeed, the generalized number of flavors can be simply expressed in function of quantities in the unbroken gauge group, irrespective of the specific symmetry breaking pattern.  In fact, only the massive gauge bosons add up to the generalized number of flavor after symmetry reduction.  Hence, for a given gauge subgroup $H$ of $G$, the ISS condition \eqref{eq:ISSC} becomes, after substituting  $N_f$ \eqref{eq:Nf_natural_def} and using identity \eqref{eq:id_casimir} [once for $N_f$ and once for $C(\h)$],
\begin{align}
\sum_R\sum_im_i\dim\pr{r_i'}C(r_i)+\cdots<\frac{3}{2}C(\h)\quad\Rightarrow\quad\xi^2\sum_RC(R)+\cdots<\frac{3}{2}\xi^2C(\g)-\cdots.\label{eq:chain}
\end{align}
The ellipses here stand for the massive gauge bosons.  From \eqref{eq:chain}, it is clear that the ISS condition for the broken group is more restrictive than the ISS condition for the unbroken group (for which the ellipses on the right-hand side are absent).

Since the relevant symmetry breaking pattern can be reached by following a specific chain of maximal subgroups, the previous result implies that the less stringent constraint on the generalized number of flavors occurs through the first breaking to maximal subgroups. Therefore, \eqref{eq:ISSC} is the only condition we consider for the remainder of the proof.

Our task is now reduced to finding all irreducible representations $R$ of $G_\text{SGUT}$ that have an acceptable Casimir, and then check if any combination of them can yield the SM content after symmetry breaking, providing that the sum of their Casimirs is still allowed.

\subsection{Ordering of Irreducible Representations}

Before looking into the allowed irreducible representations, it is convenient to introduce an ordering for the different irreducible representations.  Indeed, by ordering the irreducible representations in function of their Casimirs, it is possible to avoid computing them for all irreducible representations.  Although here the ordering must be found only for $SU(N)$ and $SO(N)$, the results below will be valid for all classical Lie algebras $\an$, $\bn$, $\cn$ and $\dn$, where $n$ is the rank.

To tackle this problem, we use basic knowledge about semi-simple Lie algebras.  The necessary concepts are quickly reviewed in Appendix~\ref{app:lie_algebras}.  In the notation employed here, $d_i^W$ stands for the $i$-th Dynkin coefficients of an irreducible representation or a root $W$ and its decomposition under the simple roots is denoted by $k_i^W$, such that they are linked by $d^W=k^WA$, with $A$ the Cartan matrix of the algebra.  Moreover, the simple roots are denoted $\alpha_i$, and they are normalized so that the largest ones have a squared size of $1$.

To begin with, we need a practical way of computing the Casimir for any irreducible representation.  As stated previously, there are multiple ways to define the Casimir $C(R)$ of an irreducible representation $R$ of $G$ with algebra $\g$.  An expression based on group theoretic arguments is \cite{Peskin1995}
\begin{align}
C(R)=\frac{\dim R}{\dim\g}C_2(R).\label{eq:casimir_formula}
\end{align}
The dimension of an irreducible representation $R$ can be found by Weyl's dimension formula \cite{cahn2014semi}, which is
\begin{align*}
\dim R=\prod_{\beta\,>\,0}\frac{\left<\beta,R+I\right>}{\left<\beta,I\right>}=\prod_{\beta\,>\,0}\frac{\sum_{i=1}^nk_i^\beta(d_i^R+1)\alpha_i^2}{\sum_{i=1}^nk_i^\beta\alpha_i^2},
\end{align*}
where the product is taken over all the positive roots $\beta$ of the algebra, and $I$ corresponds to the combination of roots that has a $1$ for each of its Dynkin coefficients.  The quadratic Casimir is given by \cite{cahn2014semi}
\begin{align}
C_2(R)=\left<R,R+2I\right>=\frac{1}{2}\sum_ik_i^R(d_i^R+2)\alpha_i^2=\frac{1}{2}\sum_{i,j}d_i^RA^{-1}_{ij}(d_j^R+2)\alpha_j^2,\label{eq:quad_casimir_formula}
\end{align}
where $A^{-1}_{ij}$ is the inverse Cartan matrix, presented in Appendix~\ref{app:lie_algebras}.  From these formulas, we can establish some ordering between the irreducible representations of an algebra.

Consider some irreducible representation $R$ of a given Lie algebra $\g$ with Dynkin coefficients $d_i^R$ and another irreducible representation $R'$ of the same algebra related to the first by $d_i^{R'}=d_i^R+\delta_{i\ell}$, with $\ell=1,2,\dots,n$, so that the only difference between $R$ and $R'$ is $d_\ell^{R'}=d_\ell^R+1$.  We then prove that
\begin{align}
C(R')>C(R).\label{eq:casimir_inequality}
\end{align}

First, we show that $\dim R'>\dim R$ since
\begin{align*}
\frac{\dim R'}{\dim R}=\prod_{\beta\,>\,0}\frac{\sum_ik_i^\beta(d_i^{R'}+1)\alpha_i^2}{\sum_ik_i^\beta(d_i^{R}+1)\alpha_i^2}=\prod_{\beta\,>\,0}\frac{\sum_ik_i^\beta(d_i^{R}+\delta_{i\ell}+1)\alpha_i^2}{\sum_ik_i^\beta(d_i^{R}+1)\alpha_i^2}>1.
\end{align*}
Indeed, the roots $\beta$ are positive and the $k_i^\beta$ are necessarily positive by definition.  Thus, we have that the numerator is equal to the denominator if $k_\ell^\beta=0$, else it is greater.  But there is always at least a root for which $k_\ell^\beta\neq0$: The $\ell$-th simple root $\alpha_\ell$ is a positive root and has $k_i^{\alpha_\ell}=\delta_{i\ell}$.  This implies the inequality $\dim R'>\dim R$.

Second, we have $C_2(R')>C_2(R)$ since
\begin{align*}
C_2(R')-C_2(R)&=\frac{1}{2}\sum_{i,j}d_i^{R'}A^{-1}_{ij}(d_j^{R'}+2)\alpha_j^2-\frac{1}{2}\sum_{i,j}d_i^R A^{-1}_{ij}(d_j^R+2)\alpha_j^2\nonumber\\
&=\frac{1}{2}\sum_{i,j}(d_i^R+\delta_{i\ell})A^{-1}_{ij}(d_j^{R}+\delta_{j\ell}+2)\alpha_j^2-\frac{1}{2}\sum_{i,j}d_i^R A^{-1}_{ij}(d_j^R+2)\alpha_j^2\nonumber\\
&=\frac{1}{2}A^{-1}_{\ell\ell}\alpha_\ell^2+\frac{1}{2}\sum_{i}[d_i^RA^{-1}_{i\ell}\alpha_\ell^2+(d_i^R + 2)A^{-1}_{\ell i}\alpha_i^2]>0,
\end{align*}
because $A^{-1}$ has only strictly positive components, regardless of the algebra, so that each term is greater or equal than zero and some terms are strictly greater than zero.  Since $\dim\g$ is the same for $R$ and $R'$, \eqref{eq:casimir_formula} implies that $C(R')>C(R)$, and this is valid for every semi-simple Lie algebra.  This result also holds for any irreducible representations $R'$ and $R$ such that $d_i^{R'}\geq d_i^R$ $\fa i$.  Indeed, in this case, $R'$ can be obtained by successively adding ones to the Dynkin coefficients of $R$, so the inequality can be applied at each step of the chain.

Of course, one only gets a partial ordering of the irreducible representations from this, since it does not allow some irreducible representations to be compared.  For example, if $R=(1,1,0,0)$, $R'=(1,2,0,0)$ and $R''=(2,1,0,0)$, then, according to these results, we have $C(R')>C(R)$ and $C(R'')>C(R)$, but the formalism does not tell us if $C(R')>C(R'')$ or $C(R')<C(R'')$.  Such an ordering will be introduced for specific groups when necessary.


\section{The No-Go Theorem}\label{sec:no-go_theorem}

Our goal is now clear, we must find the largest irreducible representations that satisfy the ISS condition \eqref{eq:ISSC}.  Unfortunately, it will be shown below that the largest irreducible representation is quite small and the few acceptable irreducible representations cannot accommodate the MSSM.

To follow with the proof, we need a more specific approach with respect to the irreducible representations, so we study theories unified with $SU(N)$ or $SO(N)$ separately.  However, the analysis will be similar in each case: We state the appropriate ISS condition and then treat irreducible representations in order of increasing Dynkin coefficients.  We find that only the irreducible representations $\one$ and $\delta_{i1}$ [and the conjugate representation $\delta_{in}$ for $SU(n+1)$] are allowed to build the SM families.  Finally, we study the possible symmetry breaking patterns and the branching rules associated with them to conclude that there is no representations that can account for the $Q_L$ quark doublets, hence completing the proof of the no-go theorem.

\subsection{\texorpdfstring{Admissible Field Content of a $SU(n+1)$-based Unified Theory}{Admissible Field Content of a SU(n+1)-based Unified Theory}}\label{SSec:G_SGUT=SU(n+1)}

We first focus on the unified group $G_\text{SGUT}=SU(N)$.  For convenience with respect to conjugate representations, we write $N=n+1$ to make the rank $n$ of the group explicit.  We keep in mind that the MSSM is of rank $4$, so we need $n$ at least equal to $5$.  We begin by finding the maximal subgroup with the largest upper bound on the generalized number of flavors.  We follow by finding an exhaustive list of the irreducible representations of $SU(n+1)$ that can be used in the original UV theory such that SUSY is broken by a metastable vacuum.

From section \ref{sec:ISS}, we only need to find the less stringent constraint on the generalized number of flavors occuring from the breaking of the SGUT gauge group to one of its maximal subgroups.  A quick analysis of possible maximal subgroups of $SU(n+1)$ determines that the weakest condition on $N_f$ corresponds to the maximal embedding $SU(n+1)\downarrow SU(n-m+1)\times SU(m)\times U(1)$ with $m>4$.  This leads to
\begin{align}
N_f<\frac{3}{2}(n-m+1),\quad m>4,\label{eq:ISS_SU_from_SU}
\end{align}
which is general for any symmetry breaking patterns and will prove to be sufficient to complete our proof.

We are now ready to construct a list of allowed irreducible representations in the UV theory.  The first irreducible representation is the trivial $\one$ with $d_i^\one=0$ $\fa i$.
This irreducible representation has $C(\one)=0$ in any group.  Then, we look at the defining representations, which we label by their Dynkin coefficients as $d_i^R=\delta_{i\ell}$ with $\ell=1,2,\dots,n$.  These constitute the building blocks of other irreducible representations according to our ordering.  For simplicity, hereafter irreducible representations will most of the time be denoted directly by their Dynkin coefficients.  Since the algebra of $SU(n+1)$ is $\an$, it is straightforward to compute the Casimir $C(R)$ for these irreducible representations from \eqref{eq:casimir_formula} and the explicit Cartan matrix presented in Appendix~\ref{app:lie_algebras}.  One has $\dim\an=n(n+2)$, $\dim\delta_{i\ell}=\binom{n+1}{\ell}$ and $C_2(\delta_{i\ell})=\frac{\ell(n+1-\ell)(n+2)}{2(n+1)}$, which leads to
\begin{align}
C(\delta_{i\ell})=\frac{1}{2}\binom{n-1}{\ell-1}.\label{eq:casimir_def_rep_SU(N)}
\end{align}
This formula is symmetric under the interchange $\ell\to n+1-\ell$ as it should since $\delta_{i,n+1-\ell}$ is the conjugate representation of $\delta_{i\ell}$.  This fact allows us to focus on $\ell\leq\floor{\frac{n+1}{2}}$ and then extend the results with the help of the symmetry.

The inequality \eqref{eq:casimir_inequality} tells us that we have $C(\delta_{i\ell})>C(\one)$, but does not allow us to compare the defining representations between themselves.  Nevertheless, with the explicit expression for their Casimirs, this can be done as follows
\begin{align}
\frac{C(\delta_{i\ell})}{C(\delta_{i,\ell-1})}=\frac{(n+1-\ell)!(\ell-2)!}{(n-\ell)!(\ell-1)!}=\frac{n+1-\ell}{\ell-1}\geq\frac{n+1}{n-1}>1.\label{eq:antisymmetric_casimir_inequality_SU(n+1)}
\end{align}
This comparison assumes $\ell\geq2$ and is thus valid only for $n\geq3$ which is verified since $n\geq5$.  Equation \eqref{eq:antisymmetric_casimir_inequality_SU(n+1)} is transformed in an inequation by replacing $\ell$ by any value $\ell\leq\floor{\frac{n+1}{2}}$.  Hence we have that the Casimir is strictly increasing with $\ell$ for $\ell=1,2,\dots,\floor{\frac{n+1}{2}}$ and then decreases symmetrically for the other values of $\ell$ by the symmetry of the Casimirs.  The Casimirs \eqref{eq:casimir_def_rep_SU(N)} of the first three defining representations are presented in the first column of Table~\ref{tab:casimir_SU(n+1)}.  According to \eqref{eq:ISS_SU_from_SU}, we have that for $\delta_{i3}$, the upper bound on the generalized number of flavors is already overcome.  Hence, inequality \eqref{eq:antisymmetric_casimir_inequality_SU(n+1)} means that only the defining representations $\delta_{i1}$ and $\delta_{i2}$ and their conjugates are allowed in the theory.  Moreover, when we consider three generations of fermions for the MSSM content, we need $3C(R)$ to not exceed the upper bound of \eqref{eq:ISS_SU_from_SU}, which furthermore reduces the MSSM field candidates to only $\delta_{i1}$ and its conjugate $\delta_{in}$.
\begin{table}[t!]
\begin{center}
    \caption{The Casimirs of some irreducible representations of $SU(n+1)$.  These can be found with the help of the formula presented in Appendix~\ref{app:lie_algebras}.  The shaded cells contain the irreducible representations that are allowed for the theory to exhibit a metastable SUSY-breaking vacuum.  The representations are ordered in columns according to the sum of their Dynkin coefficients.}
    \label{tab:casimir_SU(n+1)}
	\setlength{\arrayrulewidth}{0.6pt}
  \begin{tabular}[c]{|Sc Sc|Sc Sc|Sc Sc|} \hline
      \cellcolor{white}$R$ & \cellcolor{white}$C(R)$ & \cellcolor{white}$R$ & \cellcolor{white}$C(R)$ & \cellcolor{white}$R$ & \cellcolor{white}$C(R)$\\ \hline\hline

      \cellcolor{gray!40}$\delta_{i1}$ & \cellcolor{gray!40}$\frac{1}{2}$ & \cellcolor{gray!40}$2\delta_{i1}$ & \cellcolor{gray!40}$\frac{1}{2}(n+3)$
      & \cellcolor{white}$3\delta_{i1}$ & \cellcolor{white}$\frac{1}{4}(n+3)(n+4)$ \\

      \cellcolor{gray!40}$\delta_{i2}$ & \cellcolor{gray!40}$\frac{1}{2}(n-1)$ & \cellcolor{gray!40}$\delta_{i1} + \delta_{in}$ & \cellcolor{gray!40}$n+1$ & \cellcolor{white}$2\delta_{i1} + \delta_{in}$ & \cellcolor{white}$\frac{1}{4}(n+3)(3n+2)$\\

      \cellcolor{white}$\delta_{i3}$ & \cellcolor{white}$\frac{1}{4}(n-1)(n-2)$ & \cellcolor{white}$\delta_{i1} + \delta_{i2}$ & \cellcolor{white}$\frac{1}{2}(n^2+2n-2)$ & \cellcolor{white} & \cellcolor{white}\\

      \cellcolor{white} & \cellcolor{white} & \cellcolor{white}$2\delta_{i2}$ & \cellcolor{white}$\frac{1}{6}(n+3)(n-1)$ & \cellcolor{white} & \cellcolor{white}\\

      \cellcolor{white} & \cellcolor{white} & \cellcolor{white}$\delta_{i1} + \delta_{i,n-1}$ & \cellcolor{white}$\frac{1}{4}(n-1)(3n+4)$ & \cellcolor{white} & \cellcolor{white}\\
    \hline
  \end{tabular}
\end{center}
\end{table}
We now use inequality \eqref{eq:casimir_inequality} to find out the other possible irreducible representations.  The possible Dynkin coefficients of the representations are given by adding the Dynkin coefficients of two allowed defining representations (thus combining two among $\delta_{i1}$, $\delta_{i2}$ $\delta_{i,n-1}$ and $\delta_{in}$).  These are presented in the second column of Table~\ref{tab:casimir_SU(n+1)} along with their Casimirs, without the conjugate representations since their Casimirs are the same.  It is straightforward to verify that only the Casimirs of $2\delta_{i1}$ and $\delta_{i1}+\delta_{in}$ do not overtake the upper bound of \eqref{eq:ISS_SU_from_SU}, but three times their Casimirs do, so they cannot be used to build the MSSM families.

In the same way, we check for other irreducible representations by adding the Dynkin coefficients of the allowed representations of the second column with those of the first column, which leads to the irreducible representations of the third column of Table~\ref{tab:casimir_SU(n+1)}.  It is easy to check that they do not satisfy the ISS condition, hence, the list of allowed irreducible representations is complete.

In summary, our analysis implies that the three MSSM families can only originate from the irreducible representations $\one$, $\delta_{i1}$ and $\delta_{in}$.  Moreover, only $\delta_{i1}$, $\delta_{i2}$, $2\delta_{i1}$ and $\delta_{i1}+\delta_{in}$ and their conjugates can cause symmetry breaking by acquiring VEVs.  This fact is highlighted in Table~\ref{tab:casimir_SU(n+1)} with shaded cells.

Since the remaining steps of the proof are similar for both $SU(N)$ and $SO(N)$ fully SGUTs, we now turn our attention to unification based on $SO(N)$.

\subsection{\texorpdfstring{Admissible Field Content of a $SO(N)$-based Unified Theory}{Admissible Field Content of a SO(N)-based Unified Theory}}\label{ssec:G_SGUT=SO(N)}

We now consider a theory with $G_\text{SGUT}=SO(N)$.  The analysis is almost the same as for the $SU(N)$ case, taking into account minor changes.  As before, we first find the largest possible upper bound on $N_f$, then we obtain all irreducible representations of $SO(N)$ that can be used in the UV theory that satisfy the previous upper bound.

We have already shown in section~\ref{sec:preliminaries} that for every reduction of a symmetry group to a maximal subgroup, the upper bound on $N_f$ decreases.  Thus, to find the weakest ISS condition, we only have to find the maximal subgroup of $SO(N)$ for which the upper bound on $N_f$ is the largest.  This corresponds to the symmetry breaking pattern $SO(N)\downarrow SO(N-M)\times SO(M)$ with $M$ minimally greater than $9$ to respect the rank condition of the MSSM.  Accordingly, the ISS criterion we work with is
\begin{align}
N_f<\frac{3}{2}(N-M-2),\quad M>9.\label{eq:ISS_SO_from_SO}
\end{align}
This inequality dictates which irreducible representations of $SO(N)$ can be included in our theory.  To determine the relevant irreducible representations, we again establish an ordering between the defining representations of $SO(N)$.  The algebra of $SO(N)$ is different whether $N=2n+1$ or $N=2n$.  In the former case, the algebra is $\bn$ and the last root is different.  Hence we consider only defining representations up to $\ell\leq n-1$.  In the latter case, the algebra is $\dn$ and the last two roots are different, thus we limit ourselves to representations with $\ell\leq n-2$.  The omitted representations are studied separately afterwards.

The Casimirs for the defining representations $\delta_{i\ell}$ are the same regardless of the parity of $N$ for the values of $\ell$ considered.  We have $\dim\delta_{i\ell}=\binom{N}{\ell}$, $\dim\mathfrak{so}(N)=\frac{N(N+1)}{2}$ and $C_2(\delta_{i\ell})=\frac{\ell}{2}(N-\ell)$, leading to
\begin{align}
C(\delta_{i\ell})=\binom{N-2}{\ell-1}.\label{eq:casimir_def_rep_SO(N)}
\end{align}
As in the $SU(n+1)$ case, we compare the Casimirs of $\delta_{i\ell}$ and $\delta_{i,\ell-1}$ to provide an ordering,
\begin{align}
\frac{C(\delta_{i\ell})}{C(\delta_{i,\ell-1})}=\frac{(N-\ell)!(\ell-2)!}{(N-1-\ell)!(\ell-1)!}=\frac{N-\ell}{\ell-1}>1.\label{eq:antisymmetric_casimir_inequality_SO(N)}
\end{align}
It is easy to check that the last inequality holds for the specified values of $\ell$.

The first column of Table~\ref{tab:casimir_SO(N)} presents the Casimirs of the first three defining representations computed using \eqref{eq:casimir_def_rep_SO(N)}.  The shaded cells contain irreducible representations that do not violate the ISS condition \eqref{eq:ISS_SO_from_SO}.  We can see that for $\ell=1,2$, the representations are allowed, but not for $\ell=3$.  Moreover, only $\delta_{i1}$ is acceptable to give three generations of MSSM fermions.  Hence, according to inequality \eqref{eq:antisymmetric_casimir_inequality_SO(N)}, we do not need to consider larger values of $\ell$.
\begin{table}[t!]
\begin{center}
    \caption{The Casimirs of some irreducible representations of $SO(N)$.  The shaded cells contain the irreducible representations that are allowed for the theory to exhibit a metastable SUSY-breaking vacuum.  The representations are ordered in columns according to the sum of their Dynkin coefficients.}
    \label{tab:casimir_SO(N)}
    \setlength{\arrayrulewidth}{0.6pt}
  \begin{tabular}[c]{|Sc Sc|Sc Sc|Sc Sc|} \hline
      \cellcolor{white}$R$ & \cellcolor{white}$C(R)$ & \cellcolor{white}$R$ & \cellcolor{white}$C(R)$ & \cellcolor{white}$R$ & \cellcolor{white}$C(R)$\\ \hline\hline

      \cellcolor{gray!40}$\delta_{i1}$ & \cellcolor{gray!40}$1$
      & \cellcolor{gray!40}$2\delta_{i1}$ & \cellcolor{gray!40}$\frac{1}{2}(N+2)(N-1)$
      & \cellcolor{white}$3\delta_{i1}$ & \cellcolor{white}$\frac{1}{2}(N+1)(N+4)$\\

      \cellcolor{gray!40}$\delta_{i2}$ & \cellcolor{gray!40}$N-2$
      & \cellcolor{white}$\delta_{i1} + \delta_{i2}$ & \cellcolor{white}$(N+2)(N-2)$
      & \cellcolor{white} & \cellcolor{white}\\

      \cellcolor{white}$\delta_{i3}$ & \cellcolor{white}$\frac{1}{2}(N-3)(N-2)$
      & \cellcolor{white}$2\delta_{i2}$ & \cellcolor{white}$\frac{1}{3}(N+2)(N+1)(N-3)$
      & \cellcolor{white} & \cellcolor{white}\\
    \hline
  \end{tabular}
\end{center}
\end{table}
We still have to check the spinorial representations of $SO(N)$.  For $N=2n+1$, we had put aside $\delta_{in}$ which has Casimir $C(\delta_{in})=2^{n-3}$.  For $N=2n$, we need to verify $\ell=n-1,n$.  These two representations are conjugate to each other, with Casimir $C(\delta_{i,n-1})=C(\delta_{in})=2^{n-4}$.  One thus concludes that these representations are unusable to construct our fully SGUT.

We now use \eqref{eq:casimir_inequality} to explore more general irreducible representations.  There are four cases to consider and they are gathered in the second and third columns of Table~\ref{tab:casimir_SO(N)}.  In summary, we are left with two irreducible representations that can be used for three generations of MSSM fermions, the irreducible representations $\one$ and $\delta_{i1}$.  In addition to these representations, one can introduce one or two $\delta_{i2}$ and $2\delta_{i1}$ to build a Higgs sector, to which we now turn.

\subsection{Higgs Sector and Branching Rules}\label{ssec:higgs_sector}

We now have an exhaustive list of allowed representations for any $SU(n+1)$ or $SO(N)$ fully SGUT theory.  The last step of the proof is to show that these are not able to generate the MSSM field content.  To formally prove this statement, all symmetry breaking patterns must be investigated, and each of these patterns produces specific branching rules for the irreducible representations.  To solve this problem, we analyze the various symmetry breaking patterns that can occur from a Higgs sector built from the allowed representations.  We then compute the general branching rules for these patterns to conclude that the MSSM cannot be generated from these representations.

We begin by studying achievable symmetry breaking patterns for $SU(n+1)$ and $SO(N)$ theories.  This question was already answered in \cite{Li1974,Elias1975comment}.  The results of \cite{Li1974,Elias1975comment} are summarized in Table~\ref{tab:symmetry_breaking_SU(n+1)}.
\begin{table}[t!]
\begin{center}
    \caption{Symmetry breaking patterns from the VEV of some irreducible representations of $SU(n+1)$ or $SO(N)$.  The results are taken from \cite{Li1974,Elias1975comment}.}
    \label{tab:symmetry_breaking_SU(n+1)}
  \begin{tabular}[c]{|Sc|Sc Sc|}\hline
    $R$
    & $SU(n+1)\downarrow$
    & $SO(N)\downarrow$\\\hline\hline

    $\delta_{i1}$ or $\delta_{in}$ 
    & $SU(n)$ 
    & $SO(N)$\\ \hline

    $k[\delta_{i1}]$ or $k[\delta_{in}]$ 
    & $SU(n+1-k)$ 
    & $SO(N-k)$ \\ \hline

    $\delta_{i2}$ or $\delta_{i,n-1}$ 
    & \begin{tabular}{@{}c@{}}$SU(n)$ \\ or \\$SO(n+1)$\end{tabular}
    & \begin{tabular}{@{}c@{}}$SO(N-1)$ \\ or \\$SO(\left\lfloor \frac{N}{2} \right\rfloor)\times SO(N-\left\lfloor \frac{N}{2} \right\rfloor)$\end{tabular} \\\hline

    $2\delta_{i1}$ or $2\delta_{in}$ 
    & \begin{tabular}{@{}c@{}}$SU(n-1)\times SU(2) \times U(1)$ \\ or \\ $SO(2\left\lfloor \frac{n+1}{2} \right\rfloor+1)$ \end{tabular} 
    & \begin{tabular}{@{}c@{}}$SU(\left\lfloor \frac{N}{2} \right\rfloor)\times U(1)$ \\ or \\ $SO(N-2)\times U(1)$ \end{tabular}\\\hline

    $\delta_{i1} + \delta_{in}$ 
    & \begin{tabular}{@{}c@{}}$SU(n+1-\left\lfloor \frac{n+1}{2} \right\rfloor)\times SU(\left\lfloor \frac{n+1}{2} \right\rfloor) \times U(1)$ \\ or \\$SU(n) \times U(1)$ \end{tabular} 
    & ----- \\ \hline
  \end{tabular}
\end{center}
\end{table}
As previously stated, these patterns can be dealt with by embedding them in a chain of maximal subgroups.  For each symmetry breaking pattern found in \cite{Li1974,Elias1975comment}, we need five particular cases of maximal subgroups.  Hence, we give in Table~\ref{tab:branching_rules_SU(n+1)} the branching rules of the eight irreducible representations permitted for the two patterns when the embedding group is $SU(n+1)$.  The branching rules for the three remaining patterns, where $SO(N)$ is the embedding group, are presented in Table~\ref{tab:branching_rules_SO(N)}.  For this case, only four irreducible representations are allowed.

The branching rules are given in terms of direct products and direct sums.  For example, the branching rule for the irreducible representation $\delta_{i2}$ for the symmetry breaking pattern $SU(n+1)\downarrow SU(n-m+1)\times SU(m)$ is
\begin{align*}
\delta_{i2}^{SU(n+1)}\downarrow\pr{\delta_{i2}^{SU(n-m+1)}\times\one^{SU(m)}}\oplus\pr{\delta_{i1}^{SU(n-m+1)}\times\delta_{i1}^{SU(m)}}\oplus\pr{\one^{SU(n-m+1)}\times\delta_{i2}^{SU(m)}}.
\end{align*}
In the tables, the group index under which the representations act is suppressed to simplify the notation.
\begin{table}[t!]
\begin{center}
    \caption{Branching rules for the allowed irreducible representations for some maximal subgroups of $SU(n+1)$.  In the first column, the $U(1)$ factor of the symmetry breaking pattern is not included.  The first three representations are separated from the others to indicate that they are the only representations that can be included three times or more in the theory.}
    \label{tab:branching_rules_SU(n+1)}
  \begin{tabular}[c]{|Sc | Sc Sc|}\hline
    $R$ 
    & \ml{$SU(n+1)\downarrow$ \\ $SU(n-m+1)\times SU(m)$} 
    & \ml{$SU(n+1) \downarrow$ \\ $ SO(n+1)$}\\\hline\hline

    $\one$ 
    & $\one \times \one$ 
    & $\one$ \\ 

    $\delta_{i1}$ 
    & $\delta_{i1} \times \one \oplus \one \times \delta_{i1}$ 
    & $\delta_{i1}$ \\

    $\delta_{in}$ 
    & $\delta_{i,n-m} \times \one \oplus \one \times \delta_{i,m-1}$ 
    & $\delta_{i1}$\\ \hline 

    $\delta_{i2}$ 
    & $\delta_{i2} \times \one\oplus \one \times \delta_{i2} \oplus \delta_{i1} \times \delta_{i1} $ 
    & $\delta_{i2}$ \\

    $\delta_{i,n-1}$ 
    & $\delta_{i,n-m-1} \times \one\oplus \one \times \delta_{i,m-2} \oplus \delta_{i,n-m} \times \delta_{i,m-1} $
    & $\delta_{i2}$ \\

    $2\delta_{i1}$ 
    & $2\delta_{i1} \times \one \oplus \one \times 2\delta_{i1}  \oplus \delta_{i1} \times \delta_{i1} $
    & $2 \delta_{i1} \oplus \one$ \\

    $2\delta_{in}$ 
    & $2\delta_{i,n-m} \times \one \oplus \one \times 2\delta_{i,m-1} \oplus  \delta_{i,n-m} \times \delta_{i,m-1} $
    & $2 \delta_{i1} \oplus \one$ \\

    $\delta_{i1} + \delta_{in}$ 
    & \ml{$[\delta_{i1} + \delta_{i,n-m}] \times \one \oplus \one \times [\delta_{i1} + \delta_{i,m-1}] $\\$ \oplus\, \delta_{i1} \times \delta_{i,m-1} \oplus\, \delta_{i,n-m}\times \delta_{i1}\oplus \one \times \one$}
    & $\delta_{i1} \oplus \delta_{i2}$ \\ \hline
  \end{tabular}
\end{center}
\end{table}
\begin{table}[t!]
\begin{center}
    \caption{Branching rules for the allowed irreducible representations for some maximal subgroups of $SO(N)$.  In the last two columns, we separate the symmetry breaking pattern $SO(N)\downarrow SU(n=\floor{N/2})$ for $N$ even and odd since the rules are different in each case.  The first two representations are separated from the others to indicate that they are the only representations that can be included three times or more in the theory.}
    \label{tab:branching_rules_SO(N)}
  \begin{tabular}[c]{|Sc | Sc Sc Sc|}\hline
    $R$ 
    & \ml{$SO(N) \downarrow$ \\ $ SO(N-M)\times SO(M)$}
    & \ml{$SO(2n) \downarrow$ \\ $SU(n)$}
    & \ml{$SO(2n+1) \downarrow$ \\ $SU(n)$} \\\hline\hline

    $\one$ 
    & $\one \times \one$ 
    & $\one$
    & $\one$ \\ 

    $\delta_{i1}$ 
    & $\delta_{i1} \times \one \oplus \one \times \delta_{i1}$ 
    & $\delta_{i1} \oplus \delta_{i,n-1}$
    & $\delta_{i1} \oplus \delta_{i,n-1} \oplus \one$\\ \hline

    $\delta_{i2}$ 
    & $\delta_{i2} \times \one \oplus \delta_{i1} \times \delta_{i1} \oplus \one \times \delta_{i2}$ 
    & \ml{$\delta_{i2} \oplus \delta_{i,n-2} \oplus \one $\\$ \oplus\, [\delta_{i1} + \delta_{i,n-1}]$}
    & \ml{$[\delta_{i1} + \delta_{i,n-1}] \oplus \one \oplus \delta_{i2} \oplus \delta_{i,n-2} $\\$ \oplus\, \delta_{i1} \oplus \delta_{i,n-1}$}\\

    $2\delta_{i1}$ 
    & $2\delta_{i1} \times \one \oplus \delta_{i1} \times \delta_{i1}  \oplus \one \times 2\delta_{i1}$
    & \ml{$2 \delta_{i1} \oplus 2 \delta_{i,n-1} $\\$ \oplus\, [\delta_{i1} + \delta_{i,n-1}]$}
    & \ml{$2 \delta_{i1} \oplus 2 \delta_{i,n-1} \oplus [\delta_{i1} + \delta_{i,n-1}] $\\$ \oplus\, \delta_{i1} \oplus \delta_{i,n-1} \oplus \one$}\\\hline
  \end{tabular}
\end{center}
\end{table}

We now focus on the branching rules of the trivial \one, the fundamental $\delta_{i1}$ and the antifundamental $\delta_{in}$ representations because they are the only candidates that can contain the MSSM families.  The trivial representation $\one$ has a trivial branching rule.  In fact, since the dimension of the representations must be conserved before and after symmetry breaking, we have a direct argument to write the general rule
\begin{align}
\one^{G_\text{SGUT}}\downarrow\bigtimes_{G_i}\one^{G_i},\label{eq:branching_rule_trivial}
\end{align}
where $G_i$ is any subgroup product of $G_\text{SGUT}$.  Looking at the branching rules of $\delta_{i1}$ and $\delta_{in}$, it is not hard to convince oneself that the resulting representation is of the form of a direct sum of products of many trivial $\one$ and only one fundamental or antifundamental, at each step of the chain of maximal subgroups, regardless of the aimed subgroups and the starting group.  It can be roughly expressed as
\begin{align}
\delta_{ia}^{G_\text{SGUT}}\downarrow\bigoplus_{G_i}\pr{\delta_{ia'}^{G_i}\times\bigtimes_{\mathclap{\substack{G_j\\j\neq i}}}\one^{G_j}},\label{eq:branching_rule_fundamental_SU(n+1)}
\end{align}
where $a$ and $a'$ are used to designate whether it is the fundamental or antifundamental representation and $G_i$ are the subgroups of $SU(n+1)$ or $SO(N)$ resulting from the symmetry breaking, thus $G_i$ must be an $SU(M)$ or $SO(M)$ subgroup.  Moreover, when $G_i=SO(N_i)$, we necessarily have $a'=1$.

There is one last ambiguity we have not adressed yet: Can a fancy Higgs sector with several Higgs fields lead to another symmetry breaking pattern with branching rules different than \eqref{eq:branching_rule_fundamental_SU(n+1)}?  The short answer is no.  Indeed, we can always consider the symmetry breaking in steps with each Higgs further reducing the symmetry.  Consider, without loss of generality, that some Higgs acquires a VEV before the others.  Then the other Higgs would have to decompose according to one of the branching rules of Table~\ref{tab:branching_rules_SU(n+1)} or \ref{tab:branching_rules_SO(N)}.  But the resulting representations are all representations that we have already considered for symmetry breaking.  Thus the next Higgs in the list will produce a symmetry breaking pattern already included in Table~\ref{tab:symmetry_breaking_SU(n+1)}.  The argument can be repeated until the Higgs sector is completely exhausted, demonstrating the generality of our analysis and explaining why we considered all five symmetry breaking patterns although only the first cases in Tables~\ref{tab:branching_rules_SU(n+1)} and \ref{tab:branching_rules_SO(N)} lead to product groups of the form $G_\text{SB}\times G_\text{SM}$.

This observation was rather important to make, because if it were possible to break a $SU(N)$ group to $E_6$ in some way, then this theory would be a good candidate.  In $E_6$ unification, one fundamental $\rep{27}$ representation contains one complete generation of fermions, and these representations are in fact the only ones available to us in our framework.

It is possible to lift the requirement of the three families by looking at branching rules for irreducible representations $R$ in steps such that
\begin{align}
R\downarrow\bigoplus_im_ir_i^{G_\text{SB}'}\times r_i^{G_\text{SM}}\downarrow\bigoplus_{i,j}m_im_{ij}r_{i,j}^{G_\text{SB}}\times r_{i,j}^{G_\text{Sym}}\times r_i^{G_\text{SM}},\label{eq:branching_rule_rest}
\end{align}
where $G_\text{Sym}$ is a symmetry group of the different MSSM families and $m_{ij}$ are the appropriate multiplicities.  In this case, the three MSSM families would originate from the same irreducible representations and they would be related by some hidden symmetry group.  Thus the branching rules for the allowed irreducible representations in Tables~\ref{tab:branching_rules_SU(n+1)} and \ref{tab:branching_rules_SO(N)} which cannot be included three times could still be relevant in generating the three MSSM families.

\subsection{Proof of the No-Go Theorem}\label{ssec:proof}

We are finally ready to complete our proof.  After the analysis performed above, we are left with \eqref{eq:branching_rule_trivial}, \eqref{eq:branching_rule_fundamental_SU(n+1)} and \eqref{eq:branching_rule_rest} to build the MSSM, regardless of the unifying group or of the exact symmetry breaking pattern.  Unfortunately these representations cannot provide the MSSM quark $Q_L$ which has quantum number $(\mathbf{3},\mathbf{2})$ under $SU(3)_C$ and $SU(2)_L$.  In terms of Dynkin coefficients, this quark would need to be in $(1,0)^{SU(3)_C}\times(1)^{SU(2)_L}$, but such an arrangement never happens for the three irreducible representations in \eqref{eq:branching_rule_trivial} and \eqref{eq:branching_rule_fundamental_SU(n+1)}, \textit{i.e.} the irreducible representations that can be included at least three times.

Actually, the smallest irreducible representations that could provide such a field are the antisymmetric representations for $SU(n+1)$ and the spinorial representations for $SO(N)$.  However, we have already ruled out the spinorial representations of $SO(N)$.  The antisymmetric representations of $SU(n+1)$ would be ruled out by the MSSM family argument since they cannot be included three times.  In principle, it is still possible that several symmetry breakings could generate three MSSM families from a unique irreducible representation through a hidden MSSM family symmetry group $G_\text{Sym}$ as in \eqref{eq:branching_rule_rest}.  Looking at Tables~\ref{tab:branching_rules_SU(n+1)} and \ref{tab:branching_rules_SO(N)}, it is clear that the MSSM family symmetry group path is not possible.  Indeed, for symmetry breaking patterns leading to product groups of the form $G_\text{SB}\times G_\text{SM}$, the MSSM quark $Q_L$ can never be generated three times since the antisymmetric representation always breaks to an antisymmetric representation times the trivial representation or a fundamental representation times a fundamental representation.  Hence, this argument concludes the proof of the no-go theorem for a fully SGUT using an ISS metastable SUSY-breaking vacuum.


\section{Conclusion}\label{sec:conclusion}

In this article we introduced the idea of fully SGUTs, \textit{i.e.} SUSic grand unified theories that, upon symmetry breaking through the Higgs mechanism, decompose into a visible sector representing the MSSM and an extra sector with a gauge group responsible for SUSY breaking.  The important feature of fully SGUTs being that all gauge groups unify into a simple SGUT gauge group $G_\text{SGUT}$, including the extra sector gauge group.  Fully SGUTs are thus motivated by MSSM gauge coupling unification.

We then focused on a specific implementation of this framework where the extra gauge group breaks SUSY through the ISS mechanism.  Starting from simple SGUT gauge groups with complex representations, we then showed that the existence of the ISS SUSY-breaking minimum puts strong constraints on the matter content of the fully SGUT theories.  From the theory of Lie groups and Lie algebras, we found all the allowed irreducible representations in the fully SGUT for which the ISS SUSY-breaking minimum exists in the extra sector (assuming none of the matter field irreducible representations generated by symmetry breaking acquire a large mass).  Finally, from the possible symmetry breaking patterns induced through the Higgs mechanism, we demonstrated that none of the allowed representations contains the MSSM field content.  We thus showed that fully SGUTs with metastable SUSY breaking \textit{\`a la} ISS is forbidden.

In the specific framework studied here, one assumption that could perhaps be relaxed is symmetry breaking through the Higgs mechanism.  For example, if it were possible somehow to break $SU(27)$ to $E_6$ by a different symmetry breaking mechanism, a unified theory could maybe be built.  The remaining assumption that all irreducible representations generated by symmetry breaking are light and thus contribute to the running of the extra sector gauge coupling might also be relaxed by some clever model building.

It would also be interesting to look at other SUSY-breaking mechanisms for which the idea of fully SGUTs could be successfully implemented.  If they exist, it is likely that such models would not be generic.  However, such models could nevertheless be of interest -- after all, the SM itself is rather baroque.  For example, their phenomenology could help explain why superpartners have not yet been seen at the LHC or could address the electroweak hierarchy problem.  We hope to return to such models in future work.


\ack{
The authors would like to thank Pierre Mathieu for enlightening discussions on Lie algebras.  This work is supported in part by NSERC and FRQNT.
}


\setcounter{section}{0}
\renewcommand{\thesection}{\Alph{section}}

\section{Basics about classical Lie algebras}\label{app:lie_algebras}

We review here the relevant basics of classical Lie algebras.  First, there are four infinite families of classical Lie algebras, denoted by \an, \bn, \cn and \dn where $n$ is a positive integer called the \textit{rank} of the algebra.  They correspond respectively to the algebras of the groups $SU(n+1)$, $SO(2n+1)$, $Sp(2n)$ and $SO(2n)$.  Their structure is dictated by a set of linearly-dependant vectors of $n\in \mathds{N}$ components called the \textit{roots} that we here denote by $\beta_i$.  As a subset of these roots, there is a set of $n$ linearly-independant particular roots called the \textit{simple roots} denoted by $\alpha_i$, such that every other root $\beta$ is a linear combination of the $\alpha_i$, with either all positive integer coefficients or negative integer coefficients.  Thus, the roots can be split in two sets called positives roots and negative roots respectively.

There exists a scalar product between the roots, denoted by $\left<\beta_i,\beta_j\right>$.  Since $\beta=\sum_ik_i^\beta\alpha_i$, where the $k_i^\beta$ are constant coefficients, only the scalar product between the $\alpha_i$ is needed.  The result of all the scalar products is encoded in a matrix called the \textit{Cartan matrix} $A$ of the Lie algebra.  It is defined as
\begin{align}
A_{ij}=2\frac{\left<\alpha_i,\alpha_j\right>}{\left<\alpha_j,\alpha_j\right>},
\end{align}
and will be specified shortly for each different simple Lie algebra.

The irreducible representations of an algebra are often referred to by their dimensions, but it is easier to keep track of them by labelling them by their \textit{Dynkin coefficients}, which form a vector of $n$ positive integers.  The $n$ Dynkin coefficients of a given irreducible representation $R$ are denoted by $d^R_i$ in this paper.  Since the simple roots $\alpha_i$ form a basis for an $n$-dimensional space, they can be used to express the Dynkin coefficients of $R$.  In this basis, they will be denoted by $k_i^R$.  If we define the Dynkin coefficients of the simple roots as the rows of the Cartan matrix $A$, then we can switch between the two notations simply by writing $d_j^R=k_i^RA_{ij}$.  It is worth noting that the $k_i^R$ of an irreducible representation $R$ are all positive.  From this, we can write the scalar product between representations or roots $W$ and $W'$ as $\left<W,W'\right>=\frac{1}{2}\sum_ik_i^Wd_i^{W'}\alpha_i^2$, where $\alpha_i^2$ is the squared size of the root.  We choose here the convention where the largest squared size of the roots is normalized to $1$ so that $C(\delta_{i1})=\frac{1}{2}$ for the fundamental representation of $SU(N)$.

Here we give the Cartan matrices and their inverses for each simple Lie algebra.  Empty cells correspond to zeros. 

For \an, we have:
\begin{align}
    A = \begin{pmatrix*}[c]
     2 & -1    &       &       &      &  \\
    -1 &  2    & -1    &       &      &  \\
       & -1    &\ddots &\ddots &      &  \\
       &       &\ddots & 2     & -1   &  \\
       &       &       & -1    &  2   & -1\\
       &	   &       &	   & -1   &  2\\
\end{pmatrix*}, \qquad  A^{-1}_{ij} = \left\{ \def\arraystretch{2.2}
        \begin{array}{c l}
            \displaystyle
            \frac{i(n+1-j)}{n+1} & \hphantom{\quad} \textrm{if } i \leq j \\
            \displaystyle
            \frac{j(n+1-i)}{n+1} & \hphantom{\quad} \textrm{if } i > j.
        \end{array} \right.
\end{align}
For \bn, we have:
\begin{align}
    A = \begin{pmatrix*}[c]
     2 & -1    &       &       &      &   \\
    -1 &  2    & -1    &       &      &   \\
       & -1    &\ddots &\ddots &      &   \\
       &       &\ddots & 2     & -1   &   \\
       &       &       & -1    &  2   & -2\\
       &	   &       &	   & -1   &  2\\
\end{pmatrix*}, \qquad  A^{-1}_{ij} = \left\{ 
        \begin{array}{c l}
            i & \hphantom{\quad} \textrm{if } i \leq j,\, i \neq n \\
            j & \hphantom{\quad} \textrm{if } j < i, \, i\neq n \\
            \displaystyle
            \frac{j}{2} & \hphantom{\quad} \textrm{if } i = n.
        \end{array} \right.
\end{align}
For \cn, we have:
\begin{align}
    A = \begin{pmatrix*}[c]
     2 & -1    &       &       &      &   \\
    -1 &  2    & -1    &       &      &   \\
       & -1    &\ddots &\ddots &      &   \\
       &       &\ddots & 2     & -1   &   \\
       &       &       & -1    &  2   & -1\\
       &	   &       &	   & -2   &  2\\
\end{pmatrix*}, \qquad  A^{-1}_{ij} = \left\{ 
        \begin{array}{c l}
            i & \hphantom{\quad} \textrm{if } i \leq j,\, j \neq n \\
            j & \hphantom{\quad} \textrm{if } j < i, \, j \neq n \\
            \displaystyle
            \frac{i}{2} & \hphantom{\quad} \textrm{if } j = n.
        \end{array} \right.
\end{align}
For \dn, we have:
\begin{align}
    A = \begin{pmatrix*}[c]
     2 & -1    &       &       &      &   \\
    -1 &  2    & -1    &       &      &   \\
       & -1    &\ddots &\ddots &      &   \\
       &       &\ddots & 2     & -1   & -1\\
       &       &       & -1    &  2   & \\
       &	   &       & -1    &      &  2\\
\end{pmatrix*}, \qquad  A^{-1}_{ij} = \left\{ \def\arraystretch{1.8}
        \begin{array}{c l}
            i & \hphantom{\quad} \textrm{if } i \leq j,\, j \neq n-1,n \\
            j & \hphantom{\quad} \textrm{if } j < i, \, i \neq n,n-1 \\
            \displaystyle
            \frac{i}{2} & \hphantom{\quad} \textrm{if } j = n-1,n \,\&\, i \neq n-1,n\\
            \displaystyle \frac{j}{2}& \hphantom{\quad} \textrm{if } i = n-1,n \,\&\, j \neq n-1,n\\
            \displaystyle \frac{n}{4} & \hphantom{\quad} \textrm{if } i = j = n-1,n\\
            \displaystyle \frac{n-2}{4} & \hphantom{\quad} \textrm{if } i = n, j = n-1 \textrm{ or } i\leftrightarrow j.
        \end{array} \right.
\end{align}
For further informations about Lie algebras and their representations, we refer the reader to \cite{cahn2014semi}.

We now present the derivation of the formula used in section~\ref{SSec:G_SGUT=SU(n+1)} to compute the Casimir of a general class of irreducible representations of $SU(n+1)$.  We consider irreducible representations with Dynkin coefficients given by $d_i=p\delta_{i\ell}+q\delta_{ik}=(0,\dots,0,q,0,\dots,0,p,0,\dots,0)$ for some integers $\ell\geq k$.  To proceed, we use \eqref{eq:casimir_formula} and \eqref{eq:quad_casimir_formula}.  First we compute the quadratic Casimir,
\begin{align}
C_2(R)&=\left<R,R+2I\right>=\frac{1}{2}d_iA^{-1}_{ij}d_j'=\frac{1}{2}(p\delta_{i\ell}+q\delta_{ik})A^{-1}_{ij}(p\delta_{i\ell}+q\delta_{ik}+2)\nonumber\\
&=\frac{1}{2}[p^2A^{-1}_{\ell\ell}+pq(A^{-1}_{\ell k}+A^{-1}_{k\ell})+q^2A^{-1}_{kk}]+\sum_{j=1}^n(pA^{-1}_{\ell j}+qA^{-1}_{kj})\nonumber\\
&=\frac{1}{2N}\{N^2(p\ell+qk)+N[p(p\ell+qk)+qk(p+q)-p\ell^2-qk^2]-(p\ell+qk)^2\},\nonumber
\end{align}
where $N\equiv n+1$.  Second, we find the dimension of this irreducible representation using the hook factor technique,
\begin{align}
\dim R=\pr{\prod_{i=1}^p\frac{\binom{N+p-i}{k}}{\binom{p+q+\ell-i}{k}}\cdot\frac{\binom{N+p-k-i}{\ell-k}}{\binom{p+\ell-k-i}{\ell-k}}}\pr{\prod_{j=1}^q\frac{\binom{N+p+q-j}{k}}{\binom{q+k-j}{k}}}.
\end{align}
Finally, using $\dim\an=n(n+2)$, it is straightforward to obtain the Casimirs showed in Table~\ref{tab:casimir_SU(n+1)}.


\bibliography{SGUT}

\end{document}